# SCALING EFFECTS FOR ELECTROMAGNETIC VIBRATIONAL POWER GENERATORS

*Terence O'Donnell[1], Chitta Saha[1], Steve Beeby[2], John Tudor[2]*

[1]Tyndall National Institute, Cork, Ireland.
[2]University of Southampton, School of Electronics and Computer Science, Southampton, UK.

## ABSTRACT

This paper investigates how the power generated by electromagnetic based vibrational power generators scales with the dimension of the generator. The effects of scaling on the magnetic fields, the coil parameters and the electromagnetic damping are presented. An analysis is presented for both wire-wound coil technology and micro-fabricated coils.

## 1. INTRODUCTION

Electromagnetic principles are one means by which power can be harvested from ambient vibrations. This generally consists of using the vibrations to move a magnet relative to a coil, thus inducing a voltage in the coil defined by Faraday's law. In recent years there have been several reports of vibrational power generators based on the electromagnetic principle [1 – 3]. The sizes and structure of these generators have varied and various different power levels have been produced. However it is not clear what the limits are in terms of the power levels which can be generated from ambient vibrations using electromagnetic principles. Vibration based generators are generally aimed at the provision of power to wireless sensor systems. In such systems size is an important consideration and it is generally important that the system be miniaturised. From the application perspective it is therefore useful to understand the relationship between the size of the energy harvesting device and the power which can be generated.

This paper presents a study of how the power which can be generated by an electromagnetic based vibrational energy harvesting device, scales with the size of the device. In order to study the scaling effects there are two factors to be considered. The first is the mechanical energy in the system, which provides an upper limit on the power which can be generated. The second factor to be considered is the efficiency with which the mechanical energy can be transformed to electrical energy.

In a vibrational power generator the available mechanical energy is associated with the movement of a mass through a certain distance. Clearly the available mechanical energy will decrease with the dimensions as both the mass of the moving object and the distance moved is decreased. The electrical energy is extracted from the system by electromagnetic damping. The electromagnetic damping factor depends on flux linkage gradient, the number of coil turns, coil impedance and load impedance. Therefore the efficiency with which the electrical energy can be extracted will also decrease with size (as the magnitude of the magnetic fields decrease with size and the quality of the coils decrease). This study will investigate both effects.

## 2. GENERATOR STRUCTURE

The analysis is presented for the electromagnetic generator structure shown in figure 1 below. This structure consists of a coil sandwiched between magnets. The upper and lower magnets consist of two pairs of oppositely polarised magnets. This opposite polarity creates a flux gradient for the coil in the direction of movement, which in this case, is in the x-direction. This is a schematic representation of the type of electromagnetic generator presented in [4] and [5].

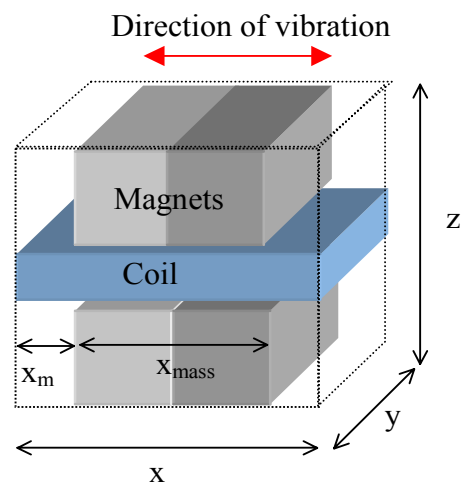

Figure 1: Schematic representation of the electromagnetic generator. In this case the magnets vibrate relative to the coil in the x-direction. The peak displacement, $x_m = (x - x_{mass})/2$.

In this structure it is assumed that the coil remains fixed while the magnets move in response to the vibration. Since the magnets generally have greater mass, m, than the coil, movement of the magnets is more beneficial than movement of the coil. By placing the coil is the gap between the magnets this structure has the advantage of creating a relatively large magnetic field in the area of the coil. The use of the two pairs of oppositely polarized magnets ensures that the coil is placed in an area of high magnetic field gradient.

For the purpose of the analysis the system is assumed to be well represented by the equations of motion of a resonant mass-spring-damper system [6]. For the purpose of this study we will assume a cubic volume, i.e. x = y = z = d where d is the dimension of the device. Note that only the magnet and coil dimensions are included in the volume. Any practical generator must also include the volume of the housing and the spring. Since the housing and the spring can be implemented in many different ways these are neglected for the present analysis. It is assumed that the spring can be implemented so as to allow the required movement of the mass at the frequency of interest.

The available mechanical energy will depend on the level of vibrations and acceleration to which the device is subjected. However the maximum displacement will be constrained by the volume of any self-contained device, i.e. the peak displacement, $x_m$, is given by the difference between the external dimension, d, and dimension of the mass, $x_{mass}$. Therefore a choice can be made to have a thin mass with a large displacement or a wide mass with a small displacement. It is of interest to find the optimum ratio of peak displacement, $x_m$, to mass dimension, $x_{mass}$, which maximises the mechanical energy. Assuming a sinusoidal vibration, and expressing the peak displacement of the mass as $x_m = (x - x_{mass})/2$, the kinetic energy of the mass can be expressed in terms of the mass dimensions and its displacement;

$$K.E = \frac{mV^2}{2} = \frac{\rho y z x_{mass} \omega^2 (x - x_{mass})^2}{8} \quad (1)$$

where V is the velocity of the mass, m, $\rho$ is the density of the mass material, $x_{mass}$, y and z are the dimension of the mass and $\omega$ is the angular frequency. By taking the derivative of this with respect to $x_{mass}$ and equating to zero it can be found that the kinetic energy is maximized when $x_m = x/3$. Thus a single magnet x-dimension is taken to be 1/6 of the overall dimension. The magnets are assumed to extend for the full y-dimension. The z-dimension of the magnet is taken as 0.4 times the overall dimension. This leaves the gap between the magnets as one fifth of the overall dimension. The coil thickness is assumed to occupy half of the gap.

The next section presents the basic equations which are used to describe the behavior of the generator.

## 3. ANALYSIS OF GENERATOR

Since the generator is assumed to be describe well by a mass-spring damper system, the basic equation of motion is given by;

$$m\frac{d^2x}{dt^2} + (D_p + D_e)\frac{dx}{dt} + kx = F_o \quad (2)$$

where $D_p$ represent parasitic damping, $D_e$ represents electromagnetic damping, k is the spring constant, and $F_o$ is the driving force, which is given by the mass times the acceleration, a. The parasitic damping represents loss mechanisms such as air damping, squeeze film effects, thermoelastic damping, and friction in the clamping. The electromagnetic damping represents the mechanism by which the electrical power is extracted from the system, i.e. the current flowing in the coil.

The typical solution for displacement given by;

$$z(t) = \frac{F_o}{\sqrt{(k - m\omega^2)^2 + (D_p + D_e)^2 \omega^2}} \quad (3)$$

The average electrical power from the generator can be expressed as;

$$P_{avg} = \frac{D_e F_0^2 \omega^2}{2[(k - m\omega^2)^2 + (D_p + D_e)^2 \omega^2]} \quad (4)$$

At resonance with $\omega = \omega_n$ this reduces to;

$$P_{resonance} = \frac{D_e F_0^2}{2(D_p + D_e)^2} \quad (5)$$

The power at resonance is maximized if $D_e = D_p$ and this maximum power, $P_{max}$, is given by;

$$P_{max} = \frac{(ma)^2}{8D_p} \quad (6)$$

Alternatively this can be expressed using the damping ratio, $\zeta_p = D_p/(2m\omega_n)$ and open circuit quality factor of the system, $Q_{oc} = 1/(2\zeta_p)$, as;

$$P_{max} = \frac{ma^2}{8\omega_n}Q_{oc} \quad (7)$$

This maximum power can only be achieved if the electromagnetic damping can be made large enough to match the parasitic damping. The electromagnetic damping factor is given by;

$$D_e = N^2 \left(\frac{d\phi}{dx}\right)^2 \frac{1}{(R_c + R_l + j\omega L_c)} \quad (8)$$

where N is the number of coil turns, $d\phi/dx$ is the average flux linkage per turn for the coil, $R_c$ is the coil resistance, $L_c$ is the coil inductance and $R_l$ is the load resistance.

The flux linkage gradient $d\phi/dx$ can be assumed to be a constant only if flux linkage varies linearly with displacement of the magnet relative to the coil. The linearity of the flux linkage can be investigated using Finite Element Analysis (FEA). Figure 2 shows the fluxlinkage vs. displacement curve obtained from a 3D FEA model of a generator with a 6 mm dimension. The magnets are assumed to be made from bulk sintered, NdFeB. As can be seen the flux linkage varies linearly over most of the displacement, but flattens out towards the end. The corresponding flux linkage and voltage vs. time plots (for an arbitrary 100 turn coil and a 100 Hz frequency).

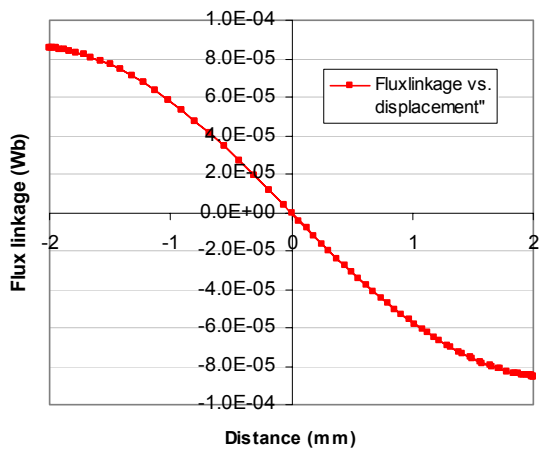

Figure 2: Flux linkage vs. displacement plot for a generator with a 6 mm outer dimension.

The flattening of the flux linkage at the outer limits of the displacement results in a distortion of the voltage waveform, which depends on the gradient of the flux linkage. The voltage is no longer perfectly sinusoidal. For the present analysis we will ignore the non-linearity and approximate the flux linkage gradient by a line fit. If the magnet size were increased relative to the coil then the flux linkage could be made linear over the displacement range, although this would deviate from the optimum dimensions for maximising kinetic energy as outlined earlier.

The other parameters which contribute to the electromagnetic damping such as coil resistance and inductance will be dependant on the number of turns in the coil. In order to maximize the power the electromagnetic damping should be chosen to be equal to the parasitic damping by choosing the number of turns for the coil and the load resistance. The next section presents an analysis of how the electromagnetic damping will scale with the dimension of the generator for the structure shown in figure 1.

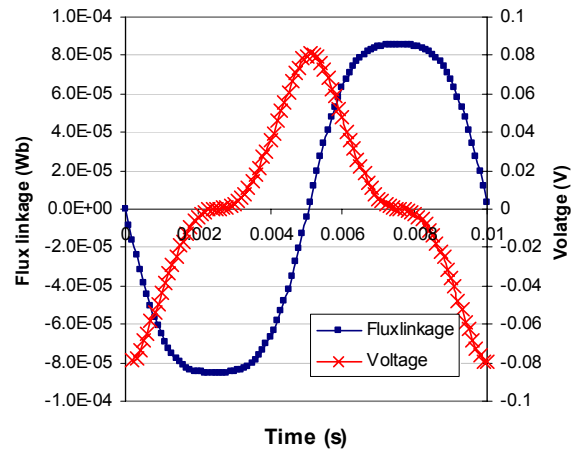

Figure 3. Flux linkage and voltage vs. time for a generator with a 6 mm outer dimension.

## 4. SCALING

For the scaling analysis, it is assumed that as the dimension is reduced, all of the relative dimensions are retained. Thus for example at a dimension of 10 mm the magnets are 4 mm in height, with a 2 mm gap between them. At a dimension of 1 mm the magnets are 0.4 mm in height with a gap of 0.2 mm between them. The scaling of the kinetic energy and the maximum power with dimension is relatively easy to determine using equations (1) and (6). However the maximum power can only be extracted if the electromagnetic damping can be controlled. As equation (8) shows, the magnitude of the electromagnetic damping depends on the magnet parameters and the coil parameters. Obviously the magnetic fields will decrease with dimension and the resistance of coils tends to increase. These effects are investigated in the following sections. Another critically important aspect of the generator is the parasitic damping. A detailed analysis of how this scales with dimension is beyond the scope of this paper, however parasitic damping is discussed in section 4.4.

### 4.1. Magnetic fields

For the electromagnetic generator the important parameter for the magnets is not the flux density but rather the flux linkage gradient. As can be seen from equation (8) the flux gradient must be maximized in order to maximize the electromagnetic damping. It would be expected that the magnitude of the field remains constant with scale reduction and that the gradient of the field would actually increase [7]. However in the present case the parameter of importance is the gradient of the flux linkage with the coil, which of course depends on the area of the coil. Therefore if the dimension of the generator is reduced by a factor, b, the flux density gradient will increase by a factor, b, and the area of the coil will decrease by a factor, $b^2$. Thus the overall effect is for the flux linkage gradient to decrease by a factor, b. This reduction can be verified by finite element analysis (FEA). The graph in figure 4 plots the flux linkage gradient obtained from 3D FEA on various generator structures with dimensions, d, varied from 1 mm to 10 mm. The graph shows a linear dependence of the gradient on the dimension.

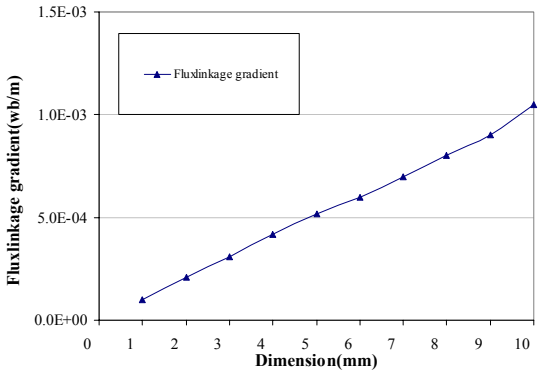

Figure 4: Variation of flux linkage gradient with generator dimension.

### 4.2. Coil Parameters

The coil parameters such as number of turns, resistance and inductance are not independent. If the coil volume is fixed then the coil resistance and inductance depends on the number of turns. If certain technology specific parameters for the coils are fixed then the resistance and inductance can be expressed in terms of the number of turns. Generally coils can be fabricated using either wire-winding or using micro-fabrication techniques.

For a wire-wound coil we assume a multilayer, circular coil with a given inner radius, $r_i$, and outer radius, $r_o$, and thickness, t. The resistance of the coil can be expressed as

$$R_c = \frac{\rho_{cu} N L_{MT}}{A_{wire}} \quad (9)$$

where ρ is the conductor resistivity, assumed to be the resistivity of copper, N is the number of turns, $L_{MT}$ is the length of the mean turn, and $A_{wire}$ is the cross sectional area of the wire. The area of the wire can be related to the overall cross sectional area of the coil, $A_{coil}$, by assuming a certain copper fill factor, $k_{cu}$, i.e. $A_{wire} = k_{cu} A_{coil} /N$. The copper fill factor depends on tightness of winding, variations in insulation thickness and winding shape. In wire wound transformers copper fill factors in the range of 0.5 to 0.6 can be achieved [8]. For a given inner and outer coil radius the coil resistance can therefore be expressed as;

$$R_c = \frac{\rho_{cu} N^2 \pi (r_o + r_i)}{k_{cu}(r_o - r_i)t} \quad (10)$$

The coil inductance can also be expressed as a function of the number of turns and the coil geometry however for the present analysis the inductance is neglected as the resistive impedance of the coil is always significantly larger than the inductive impedance at 1 kHz. It can be seen from (10) that for a constant number of turns the coil resistance is proportional to the inverse of the scaling factor provided that the copper fill factor does not depend on the scaling, (an assumption which would merit further investigation).

For a micro-fabricated coil, the resistance is more conveniently expressed in terms of the number of turns, the turn width, w, spacing, s, and thickness, t, and the coil outer, $d_o$, and inner dimensions, $d_i$. If it is assumed that the tracking spacing, track width and thickness are equal, then the coil resistance for a single layer, microfabricated coil can be expressed as;

$$R_c = \frac{8\rho_{cu}(d_o + d_i)}{(d_o - d_i)^2}(4N^3 - 4N^2 + N) \quad (11)$$

For large N, the $N^2$ and N terms can be neglected. The dependence of the coil resistance on the cube of the number of turns arises in this case, from the fact that the turn cross-sectional area decreases with the number of turns squared, i.e. both track width and thickness decrease with N. Again we can see from (11) that for a constant number of turns the coil resistance is proportional to the inverse of the scaling factor.

The number of turns in any given area can be increased only to the extent which the particular technology allows. Thus in the case of wire winding the main constraint is the minimum wire diameter, which may be as low as 12 μm. For a microfabricated coil the limitation is the minimum turn width or spacing. With advanced processes this can be submicron, although 1 μm is a reasonable practical limit for relatively standard photolithography.

### 4.3 Electromagnetic damping

In section 4.1 it was shown that the flux linkage gradient scales linearly with dimension. Equation (10) and (11) also show that the coil resistance has an inverse dependence on scaling factor. This would therefore suggest that the EM damping would have a dependence on the cube of the scale factor. However this can be compensated for to some extent, by increasing the number of turns and decreasing the load resistance as the EM damping also depends on these. However the extent to which this can be done depends on the relationship between the coil resistance and the number of turns. If the relationship in equation (10) holds for a wire wound coil then the EM damping factor can be increased by increasing the number of turns, because the coil resistance only increases at the same rate as the number of turns squared ($R_c \propto N^2$). However if the relationship in (11) holds for a micro-fabricated coil then the coil resistance increases at a faster rate than the number of turns squared (i.e. $R_c \propto N^3$) so that in this case an increase in the number of turns actually reduces the EM damping.

It is therefore the case that the EM damping scales differently with dimension depending on the relationship between coil resistance and number of turns. For a wire wound coil the EM damping can depend on scale factor squared, but for a micro fabricated coil it can depend on the scale factor cubed.

### 4.4 Parasitic damping

The parasitic damping occurs due to effects such as air damping, squeeze film damping, clamping friction for the beam, and thermoelastic damping. There have been considerable efforts to find analytical expressions for these various damping mechanisms particularly for Silicon based MEMS devices such as resonators [9][10]. These damping effects cannot be estimated without defining the beam dimensions and material which has not been addressed in this paper. However as will be shown in the next section unloaded (no EM damping) quality factors of tens of thousands are required in order to get the maximum energy from the structures considered here. Quality factors of this magnitude have been reported for Silicon based beam structures [11], however the quality factor obtained for fabricated generators have only been of the order of several hundreds [12]. In the analysis presented in the next section both levels of parasitic damping are considered. It is clear that a more thorough understanding of the sources of parasitic damping is required in order to fully optimize the generators.

### 5. DISCUSSION OF SCALING EFFECTS

In order to investigate the effects of scaling on the power which can be generated from an electromagnetic generator, we assume a vibration frequency of 1 kHz and an acceleration level of 9.81 m/s². The magnets are assumed to consist of sintered NdFeB. Using equation (6) it is relatively straightforward to plot the available the maximum electrical energy vs. the generator dimension assuming an open circuit quality factors. Since the power depends on the square of the mass then the power will scale with the sixth power of the scaling factor for a constant unloaded Q-factor. However the real scaling will depend on how the parasitic damping scales with dimension. However equation (6) assumes that the EM damping can be made equal to the parasitic damping. This can only be achieved if the parasitic damping is low enough. If the parasitic damping is very high then the situation can arise where the EM damping cannot be increased to equal the parasitic damping. This leads to two different strategies for maximizing the electrical power from the generator. Where $D_p$ is comparable to $D_e$ then power is maximized when load resistance and number of turns are chosen so as that $D_e$ is made equal to $D_p$. However when $D_p \gg D_e$ power to the load is maximized when the load and coil resistance are equal (the usual maximum power transfer condition).

Figure 5 below plots the maximum electrical power and the power delivered to the load for generator dimensions from 1 mm to 10 mm for both wire-wound coils and micro-fabricated coils. Curves are plotted for unloaded Q-factors in the range of 2000 – 30,000 which allow an unloaded displacement of twice the maximum displacement and for an unloaded Q-factor of 300, which is more typical of what has been measured on generators fabricated to date. For the large unloaded Q-factors, the characteristics of the wire-wound coils allow the EM damping to be made equal to the parasitic damping for all but the smallest dimensions. For small dimensions the number of turns is limited by the minimum wire diameter which is taken to 12 μm. For the micro-fabricated coils the EM damping cannot be made equal to the parasitic damping because of the high coil resistance.

For the unloaded Q-factor of 300 the parasitic damping is typically an order of magnitude larger than the maximum EM damping achievable even with the wire wound coil. In this situation the displacement is much less than the maximum allowed and is controlled by the parasitic damping, i.e. changes in the EM damping have little effect on the displacement. In this case the load resistance is set equal to the coil resistance in order to achieve maximum power transfer to the load.

The graph shows that if high unloaded Q-factors can be obtained then the theoretical maximum power can be extracted from the generator using a wire-wound coil. This is because the high EM damping can be achieved and little power is lost in the coil due to the resistance. However only a fraction of the theoretical maximum energy can be extracted using micro-fabricated coil due to the higher resistance which has the effect of limiting the EM damping and also increasing loss in the coil.

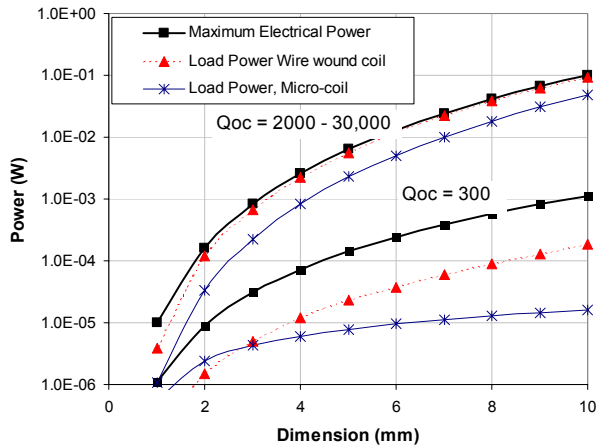

Figure 5 Maximum power and power delivered to the load for generator dimensions from 1 mm to 10 mm.

It is interesting to not that at small dimensions the power extracted using a micro-coil can exceed that extracted using a wire-wound coil due to the limitations in minimum wire diameter. The load voltages for the generators are plotted in figure 6 below.

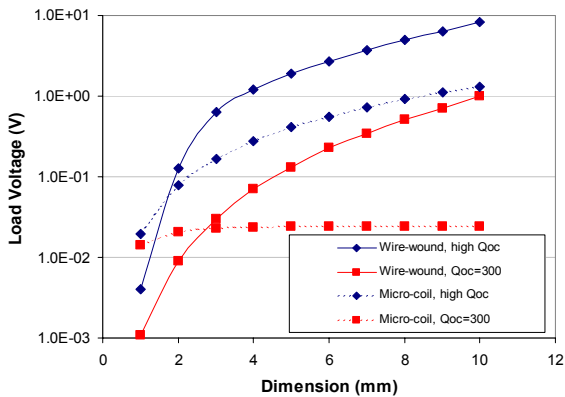

Figure 6. Load voltages for generator dimensions from 1 mm to 10 mm.

It can be seen from this that the load voltages obtained from the micro-coils are unacceptably low if the open circuit quality factor of the generator is low.

## 6. CONCLUSIONS

With high enough mechanical Q factor the performance of the generator is not limited if wire wound coils are used, i.e. the optimum condition for maximum power generation can be achieved. In this case the power scales approximately with the fourth power of the dimension. However the use of micro-fabricated coils tend to limit the generator performance due to the large coil resistance. However micro-fabricated coils will have better performance than wire-wound coils at dimensions less than 2 mm. From measured results on fabricated generators it would appear that the mechanical Q is a limiting factor. The source of is an important area for further investigation.

## 7. ACKNOWLEDGEMENTS

The authors wish to acknowledge fund for this work under the European Union Framework 6 STEP project VIBES, project reference 507911.